\documentclass{vgtc}
\usepackage[utf8]{inputenc}
\usepackage{xcolor}
\usepackage{pgfplots}
\usepackage{subfig}
\usepackage{booktabs}
\usepackage{caption}
\ifpdf
  \pdfoutput=1\relax                   
  \usepackage{graphicx}                
  \DeclareGraphicsExtensions{.pdf,.png,.jpg,.jpeg} 
\else
  \usepackage{graphicx}                
  \DeclareGraphicsExtensions{.eps}     
\fi%

\graphicspath{{figures/}{pictures/}{images/}{./}} 

\usepackage{microtype}                 
\PassOptionsToPackage{warn}{textcomp}  
\usepackage{textcomp}                  
\usepackage{mathptmx}                  
\usepackage{times}                     
\usepackage{cite}                      
\usepackage{tabu}                      
\usepackage{booktabs}                  

\onlineid{0}

\vgtccategory{Research}

\vgtcinsertpkg

\newcommand{\code}[1]{\texttt{#1}}

\title{Minority Report: A Graph Network Oracle for In Situ Visualization}
\author{Krishna Kumar \thanks{e-mail: krishnak@utexas.edu}\\ %
        \scriptsize Cockrell School of Engineering, UT Austin%
\and Paul Navratil \thanks{e-mail: pnav@tacc.utexas.edu}\\ %
        \scriptsize TACC, UT Austin%
\and Andrew Solis  \thanks{e-mail: asolis@tacc.utexas.edu}\\ %
        \scriptsize TACC, UT Austin%
\and Joseph Vantassel  \thanks{e-mail: jvantassel@tacc.utexas.edu}\\ %
        \scriptsize TACC, UT Austin}

\date{June 2022}

\teaser{
  \centering
  \includegraphics[width=\linewidth]{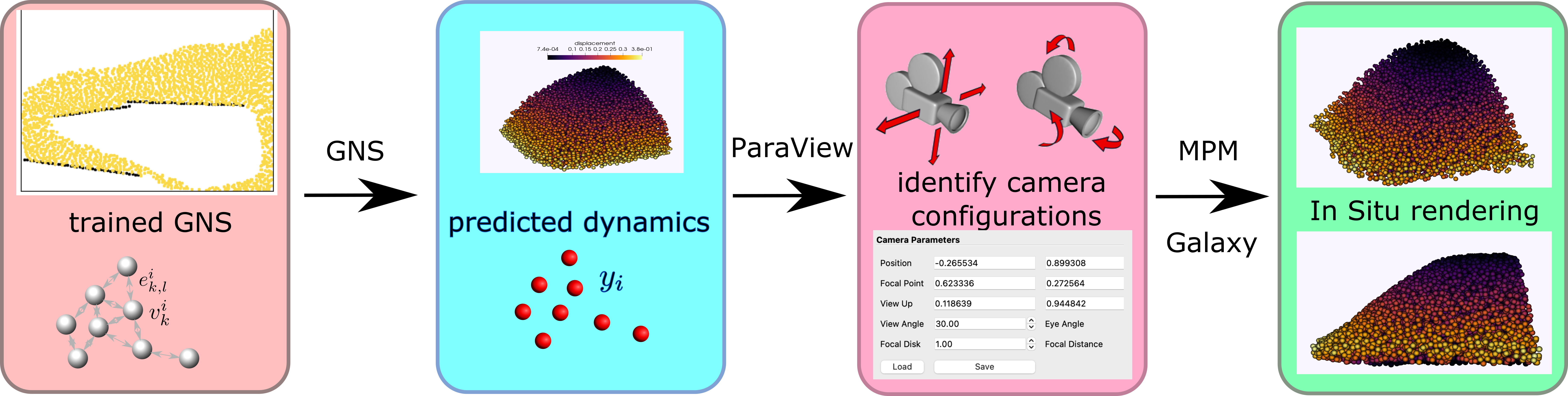}
  \caption{Overview of GNS-informed in situ visualization. (a) A GNS is trained on granular flow trajectories, (b) the trained GNS predicts the dynamics of the domain, (c) human-in-loop to identify appropriate camera configurations and select relevant features to render, and (d) set-up and run in situ visualization of CB-Geo MPM simulation of granular flow with TACC Galaxy.}
  \label{fig:teaser}
}

\abstract{In situ visualization techniques are hampered by a lack of foresight: crucial simulation phenomena can be missed due to a poor sampling rate or insufficient detail at critical timesteps. Keeping a human in the loop is impractical, and defining statistical triggers can be difficult. This paper demonstrates the potential for using a machine-learning-based simulation surrogate as an oracle to identify expected critical regions of a large-scale simulation. These critical regions are used to drive the in situ analysis, providing greater data fidelity and analysis resolution with an equivalent I/O budget to a traditional in situ framework. We develop a distributed asynchronous in situ visualization by integrating TACC Galaxy with CB-Geo MPM for material point simulation of granular flows. We employ a PyTorch-based 3D Graph Network Simulator (GNS) trained on granular flow problems as an oracle to predict the dynamics of granular flows. Critical regions of interests are manually tagged in GNS for in situ rendering in MPM.%
} 

\CCScatlist{
  \CCScatTwelve{Human-centered computing}{Visu\-al\-iza\-tion}{Visu\-al\-iza\-tion techniques}{Treemaps};
  \CCScatTwelve{Human-centered computing}{Visu\-al\-iza\-tion}{Visualization design and evaluation methods}{}
}

\begin{document}

\firstsection{Introduction}

\maketitle

In situ visualization only approximates a traditional post hoc analysis workflow. This approximation takes one of three modes: (1) an analyst interacts directly with a visualization interface connected to the running simulation (``human-in-the-loop"); (2) pre-scripted analyses are run against the simulation, either at a scheduled interval or when triggered by simulation state (``scripted"); or (3) simulation data is decimated (in time, space, data elements, or some combination) and saved for post hoc analysis (``decimation"). Each of these modes risks missing crucial simulation evolution by omitting critical timesteps, inability to express vital states to trigger analysis, or decimation data loss. More is needed to elevate the analytical power of in situ visualization to match post hoc visualization, where the data evolution can be rewound and replayed with different analysis foci.

In this paper, we propose using a machine learning (ML)-based simulation surrogate as an oracle to determine critical regions for in situ analysis (see fig.~\ref{fig:teaser}). In our method, the ML surrogate provides interactive post hoc analysis capabilities, allowing the analyst to identify simulation parameters ranges, critical timesteps, and promising data regions using a less powerful computing platform (e.g., a laptop rather than a large distributed HPC cluster).  These observations are used to target traditional in situ analysis methods to generate the most relevant results with minimal excess generation. Our method saves significant computation and I/O cost over traditional in situ methods through precise application of visualization methods in time and space (i.e., rather than a parameter sweep). We demonstrate these savings on a granular column collapse simulation and a graph network simulation surrogate.

Our work makes the following contributions: 
\begin{itemize}
    \item a dynamic graph neural network simulation surrogate to predict the roll-out of particle trajectories;
    \item a method leveraging the simulation surrogate to harvest metadata and identify parameters and features to target full-scale in situ analysis; and
    \item a demonstration of our surrogate-informed in situ visualization method on a large-scale column collapse problem, where emergent phenomena cannot be easily expressed as a closed form solution.
\end{itemize}

\noindent Watch our overview video~\url{https://youtu.be/j5qFD8lrt74}.

\section{Related Work}
\label{sec:related}

This section places our current work in the context of other in situ approaches, particularly those that leverage machine learning and those that dynamically respond to data evolution. We refer readers to Childs et~al.~\cite{childs20insitu} and Childs, Bennett and Garth~\cite{childs22insitu} for broader discussions of recent in situ work. Our technique is similar in spirit to a metadata storage in situ data product, as characterized by Patchett and Ahrens~\cite{patchett18optimizing}, in that the observations collected from the ML surrogate provide the metadata used to focus full-scale in situ computations.

\subsection{Machine Learning Surrogates for In Situ}

Machine learning surrogates are gaining popularity for in situ applications since they can produce plausible output much faster than full-scale computation. GNN-Surrogate~\cite{shi22gnn} uses a convolutional graph neural network (GNN) simulation surrogate to model a fixed-mesh ocean simulation at several fixed levels of detail in order to facilitate parameter space exploration. While our work also uses a GNN, the modeled granular column collapse simulation evolution can (and typically does) induce large structural deformations such that the GNN must predict particle dynamics and dynamically update the graph structure. We leverage message passing GNN to achieve dynamic evolution capabilities. We also use the ML surrogate to optimize full-scale traditional in situ analysis rather than operating on the ML surrogate output itself, which enables our method to capture human-tagged phenomena that is difficult to capture via an automated method.

Other recent applications of ML surrogates train on visualization outputs (rather than the simulation data directly) for feature detection~\cite{dutta18insitu} and to explore parameter space~\cite{he20insitunet}. In contrast, our work uses an ML surrogate of the simulation itself to determine metadata for full-scale in situ analysis. 

\subsection{Data-driven In Situ Techniques}
Some in situ techniques leverage automated data queries and data condition sensing to guide their operation and thus focus their attention on critical data regions and timesteps. These can take the form of feature detection for analysis (e.g., \cite{dutta18insitu}), preserving detail under data decimation  (e.g., \cite{wang17sampling,biswas2018situ,biswas21sampling}), or invoking analysis methods only when certain data conditions are met (e.g., \cite{bennett2016trigger,larsen2018flexible,lawson21insitu}).

\section{Methodology}
\label{sec:method}

This section describes our dynamic graph GNN surrogate (GNS) for granular flow simulations using the Material Point Method (MPM)~\cite{kumar2019scalable,soga2016trends} and our method for using GNS to gather metadata for full-scale, full-resolution in situ visualization within MPM using Galaxy~\cite{abram18galaxy}. We refer interested readers to Abram et~al.~\cite{abram22insitu} for details of the in situ implementation.

\begin{figure*}[!t]
    \centering
    \includegraphics[width=\textwidth]{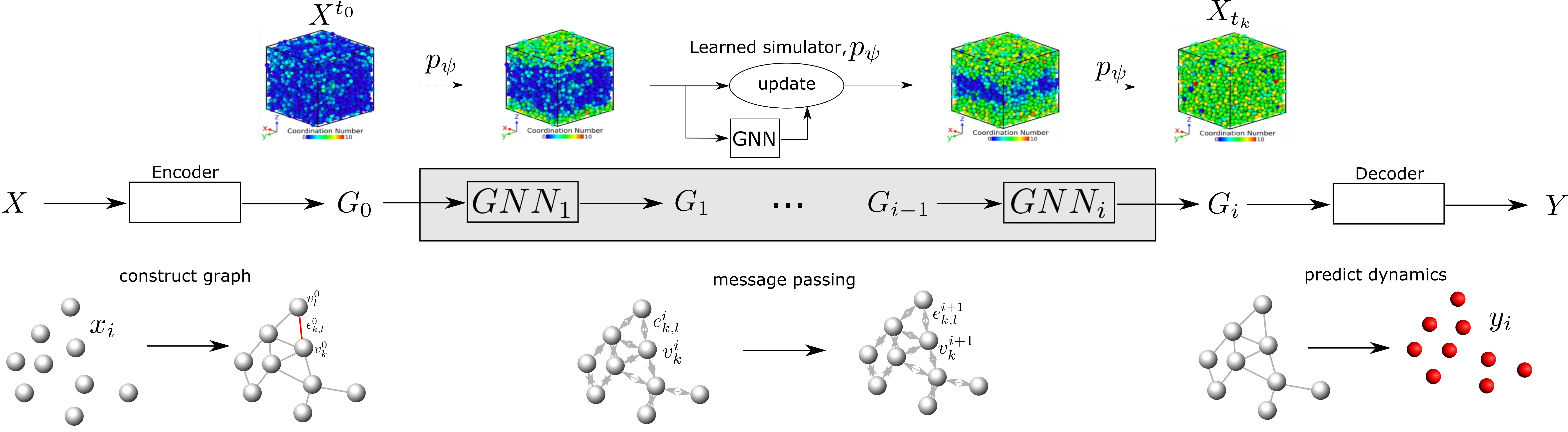}
    \label{fig:gnn}
\end{figure*}
\subsection{GNS for Large-Deformation Granular Flows}
Graph Neural Networks (GNNs)~\cite{scarselli2008graph} are state-of-the-art Geometric Deep Learning algorithms operating on graphs to represent rich relational information and local vertex features.  Graph Networks are effective in learning the interaction dynamics.  A graph network maps an input graph to an output graph with the same structure but potentially different vertex, edge, and global feature attributes.  Fig.~\ref{fig:gnn} shows an overview of the GNN learning to simulate n-body dynamics.  The graph network spans the physical domain with vertices representing an individual or a collection of particles, and the edges connecting the vertices represent the local interaction between particles or clusters of particles.  The GNN learns the dynamics, such as momentum and energy exchange, through a form of messages passing~\cite{gilmer2017neural}, where latent information propagates between vertices via the graph edges.  GNN has three components: (a) Encoder, which embeds particle information to a latent graph, the edges are learned functions; (b) Processor, which allows data propagation and computes the nodal interactions across steps; and (c) Decoder, which extracts the relevant dynamics (e.g., particle acceleration) from the graph.

GNN learns to predict the particle dynamics through message passing~\cite{sanchez2020learning}.  The GNN edge messages  ($e^\prime_k \leftarrow \phi^e(e_k, v_{r_k}, v_{s_k}, u)$) are a learned linear combination of the true forces.  The edge messages are aggregated at every node exploiting the principle of superposition $\bar{e_i^\prime} \leftarrow \sum_{r_k = i} e_i^\prime$.  The vertex then encodes the connected edge features and its local features using a neural network: $v_i^\prime \leftarrow \phi^v (\bar{e_i}, v_i, u)$.  The GNS implementation uses semi-implicit Euler integration to update the next state of the particles based on the predicted accelerations at the vertices.  We introduce physics-inspired simple inductive biases, such as an inertial frame that allows learning algorithms to prioritize one solution (constant gravitational acceleration) over another, reducing learning time.  We train the GNS model on small-scale granular collapse and collision problems with 1000 particles for 20 Million steps on NVIDIA A100 GPUs.  The trained model then accurately predicts (within 5\% of error compared to MPM simulations) the granular flow on an inclined plane with 10x the number of particles during its training.  GNN trained on trajectory data is generalizable to predict particle kinematics in complex boundary conditions not seen during training.  We developed an open-source PyTorch GNN simulator that can successfully predict the dynamics of fluid and particulate systems~\cite{Kumar_Graph_Network_Simulator_2022}.  The GNN simulator is scalable to 100,000 vertices and more than one million edges.

\subsection{GNS-driven In Situ Analysis}

We trained a GNS simulator on 30 different trajectory data for 5 million steps.  We then simulate the 3D granular column collapse experiment, which captures the dynamics of landslides.  The GNS predicts the rollout, i.e., the runout dynamics of a small-scale 15k particles granular column collapse simulation.  GNS predicts the entire runout process from initiation to collapse for the duration of one second with a time step of 0.0025~s.  The GNS rollout prediction runs on an NVIDIA GPU A100 node on  Lonestar6~\cite{lonestar6} at the Texas Advanced Computing Center (TACC).  The GNS rollouts are written as \code{vtp} files for post-processing in ParaView~\cite{ayachit15paraview}.  The GNS only takes $\sim 5$s to predict the entire rollout.  We write 500 \code{vtp} files at every timestep in the GNS prediction.  We use ParaView rendering of GNS simulations to identify feature sets and camera views for in situ visualizations. Fig.~\ref{fig:mpm-gcn-results} shows the three views (camera positions and view ports) chosen from GNS for in situ visualization of displacements using MPM. We also predict the displacement magnitude ranges (0 to 0.38 m) to color-code our in situ rendering of particles.

Once we identified the appropriate feature sets from ParaView rendering of the GNS results, we set the parameters in Galaxy to perform in situ visualization. Fig.~\ref{fig:mpm-gxy-results} shows the in situ MPM rendering for the three pre-chosen view ports from GNS. The GNS and MPM results agree reasonably well. We do notice some differences in the periphery of the simulation, which is expected as the GNS is only trained on 5 million steps. Typically, a more accurate simulation representation would require training on 20 million steps~\cite{sanchez2020learning}.

TACC Frontera \cite{stanzione20frontera} was used for running the in situ experiments. We utilize a single node with 56 cores on two sockets with a clock rate of 2.7 GHz nominal, and 192 GB DDR4 RAM.  A different run was performed for varying camera positions.  Each run output an image for every 20 timesteps, similar to the settings used for MPM.  A single galaxy process receives data from the MPM simulation to create in situ visualizations.  MPM runs in parallel across 26 cores.  Galaxy records the time it takes to receive data from MPM, including re-partitioning, so that partitions do not overlap.  The time for setting up the acceleration structures to raytrace the particles is also recorded, along with the total rendering time. 

We also performed simulations of 100k and 1 million particles using in situ MPM. We ran the 100k particles simulations on Stampede2 with two nodes and 70 total MPM processes for 5000 timesteps. For the 1 million point experiment, we used the large-memory (2TB) nodes on TACC Frontera. These nodes have 112 cores on four sockets, a clock rate of 2.7GHz nominal, and a 2.1 TB NVDIMM memory. The 1 million in situ renderings were run for 550 timesteps and used 75 MPM processes. In both cases, a single Galaxy process receives and renders the data. 

\section{Results}
\label{sec:results}

\begin{figure}[htb]
\centering
\resizebox{0.45\textwidth}{!}{%
\begin{tikzpicture}
\begin{axis}[
	legend style={at={(0.5,-0.1)},
	anchor=north,legend columns=-1},
	ybar,
	bar width = 4.5,
	xtick={0, 1, 2, 3, 4, 5, 6},
	ylabel = Time (s),
	xticklabels={run0, run0\_1, run1, run2, run3, run4, run5},
	scaled y ticks= base 10:2
]
\addplot 
	coordinates {(0,0.052)(1,0.052)(2,0.052)(3,0.052)(4,0.052)(5,0.052)(6,0.052)};

\addplot 
	coordinates {(0, 0.012)(1, 0.013)(2,0.012)(3,0.012)(4,0.012)(5,0.012)(6,0.013)};

\addplot 
	coordinates {(0,0.163)(1,0.159)(2,0.158)(3,0.162)(4,0.166)(5,0.157)(6,0.165)};

\legend{Receive,Setup,Render}
\end{axis}
\end{tikzpicture}
}
\caption{Graph of time for each in situ step. The breakdown shows the time spent for: Galaxy receiving data, setting up for rendering, and the rendering time for 15k particles simulation.}
\label{fig:insituruntimes}
\end{figure}
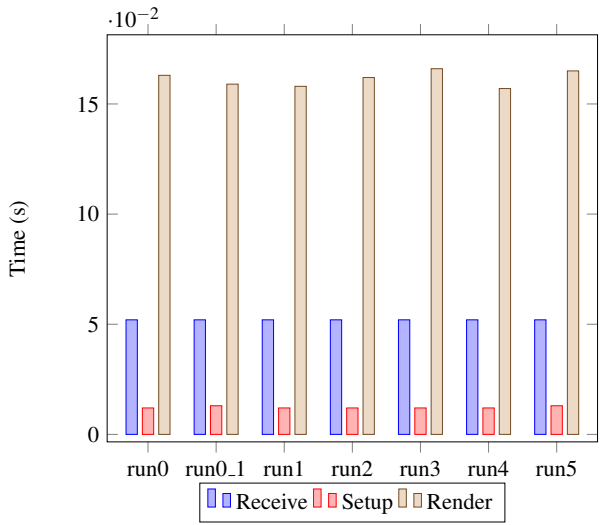

Detailed metrics of the in situ MPM with Galaxy runs for 13k particles are shown in the following images and tables. Fig.~\ref{fig:insituruntimes} shows the time for each respective in situ run. We can see that each run takes relatively the same time for receiving the data from MPM to Galaxy, setting up, and rendering. The receiving runtimes were, on average, around 0.052 seconds, with the standard deviation irrelevant. The setup time was the fastest, with 0.012 seconds and a standard deviation between 0.001 and 0.002. The rendering time was the most time-consuming ranging from 0.159 seconds to 0.166 seconds. The standard deviation ranges from 0.001 to 0.005, respectively. Table~\ref{tab:totalruntimes} shows the total runtimes for each in situ run, with the fastest being 6 minutes 51 seconds and the slowest being 10 minutes 11 seconds.

 \begin{figure}[!ht]
    \subfloat[side-view  \label{subfig-gcn-side}]{%
      \includegraphics[width=0.2\textwidth]{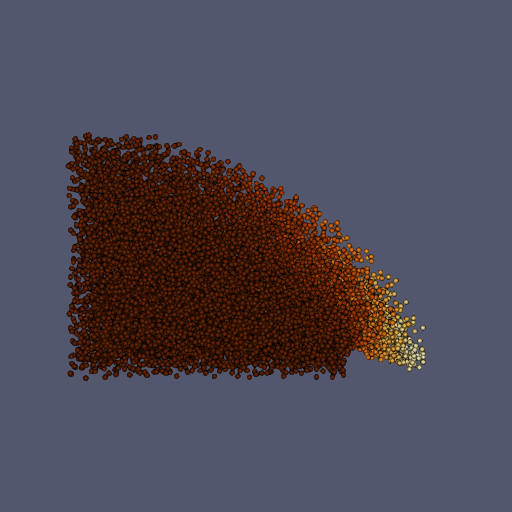}
    }
    \hfill
    \subfloat[top view  \label{subfig-gcn-top}]{%
      \includegraphics[width=0.2\textwidth]{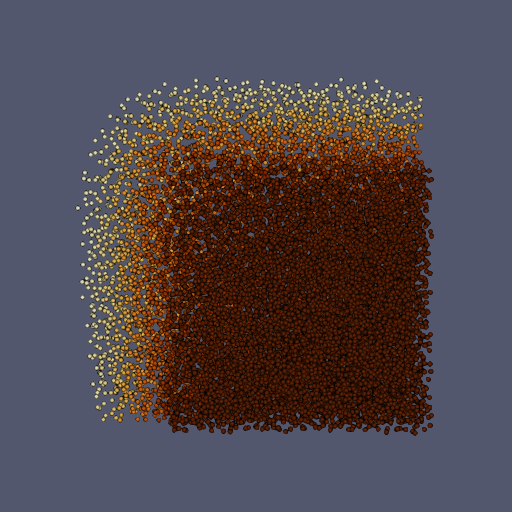}
    }
    
    \centering
    \subfloat[aerial view  \label{subfig-gcn-aerial}]{%
      \includegraphics[width=0.2\textwidth]{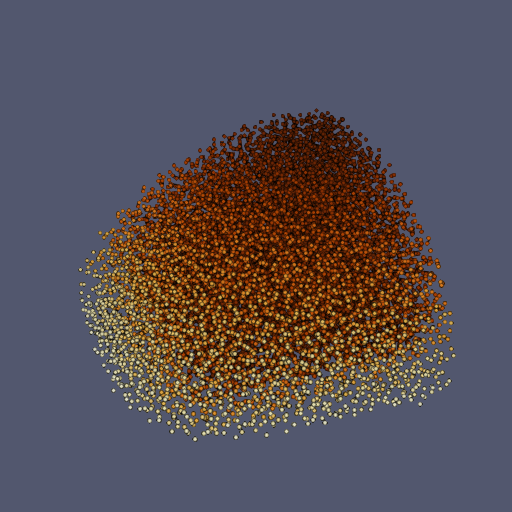}
    }
    \caption{Rendered views using vtk outputs of GNS in ParaView.}
    \label{fig:mpm-gcn-results}
 \end{figure}

\begin{figure}[!ht]
    \subfloat[side-view  \label{subfig-gxy-side}]{%
      \includegraphics[width=0.2\textwidth]{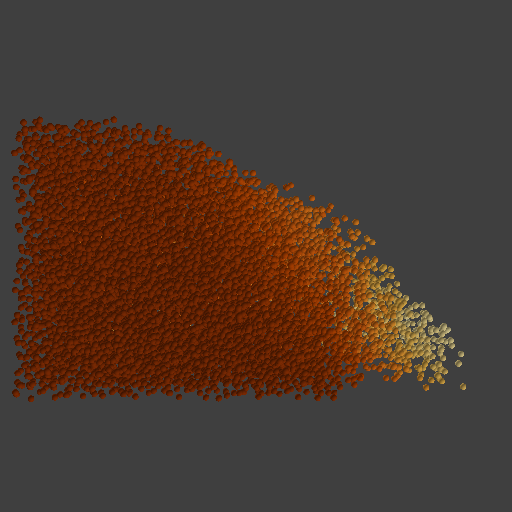}
    }
    \hfill
    \subfloat[top view \label{subfig-gxy-top}]{%
      \includegraphics[width=0.2\textwidth]{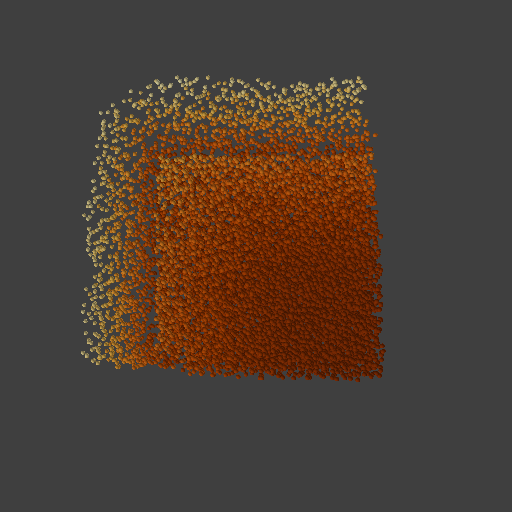}
    }
    
    \centering
    \subfloat[aerial view  \label{subfig-gxy-aerial}]{%
      \includegraphics[width=0.2\textwidth]{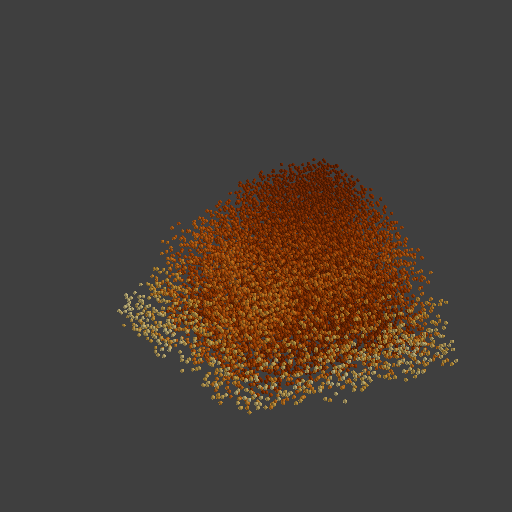}
    }
    \caption{Rendered views from in situ MPM runs with Galaxy.}
    \label{fig:mpm-gxy-results}
 \end{figure}

\begin{table}[h!]
    \centering
    \caption{Total runtime for each in situ visualization with MPM for 13k particles.}
    \label{tab:totalruntimes}
    \begin{tabular}{l c} 
     \toprule
     Run & Runtime \\ 
     \midrule
     run\_0 & 423 s\\ 
     run\_0\_1 & 511 s \\ 
     run\_1 & 504 s\\ 
     run\_2 & 486 s\\ 
     run\_3 & 420 s \\ 
     run\_4 & 412 s \\ 
     run\_5 & 612 s \\ 
     \bottomrule
    \end{tabular}
\end{table}

Figures \ref{fig:mpm-gcn-results} and \ref{fig:mpm-gxy-results} show the displacement fields from the GNS run and in situ MPM run with Galaxy. Each capture was taken around relatively same timesteps. The GNS predictions use a larger timestep of 0.0025 s compared to the MPM simulations, which uses 1E-4 s. The GNS runs were computed and then loaded into ParaView. From ParaView the camera view was changed to the desired location, with a colormap applied to the displacement of the particles. The camera positions and colormap were then transferred to be used by Galaxy for the in situ runs. Galaxy renderer used the same colormap scheme and camera angles. During the initial stages of the collapse, the side view (figs~\ref{fig:mpm-gcn-results}a and\ref{fig:mpm-gxy-results}a) are critical in capturing the shear localization. This localized failure initiation feature is not captured in the top view (fig. \ref{fig:mpm-gxy-results}b) or the aerial view (fig. \ref{fig:mpm-gxy-results}b). However, these top and aerial views are important to show the extent of runout in the final stages. Although we captured all three views in this smaller example, we could choose to render only the side-view during the inital stages of collapse (up to 0.015 s or 1500 timesteps) and then render the top and aerial views beyond 1500 steps till end of simulation. 

GNS requires a significant amount of pretraining; however, GNS simulations are generalizable to different boundary conditions, geometric configurations, and scales of the problem.  GNS scales easily to systems 10x larger than the training set.   Hence, the additional cost of running a GNS-informed in situ visualizations only increases runtime by less than 1\% for small-scale simulations (15,000 particles).  This increase in runtime cost for the GNS-informed in situ visualization is almost negligible in the case of large-scale problems (100,000 and 1 million particles), where GNS runtime cost remains constant and the MPM simulation time and data transfer increases considerably.

\begin{figure}
    \centering
    \includegraphics[width=\linewidth]{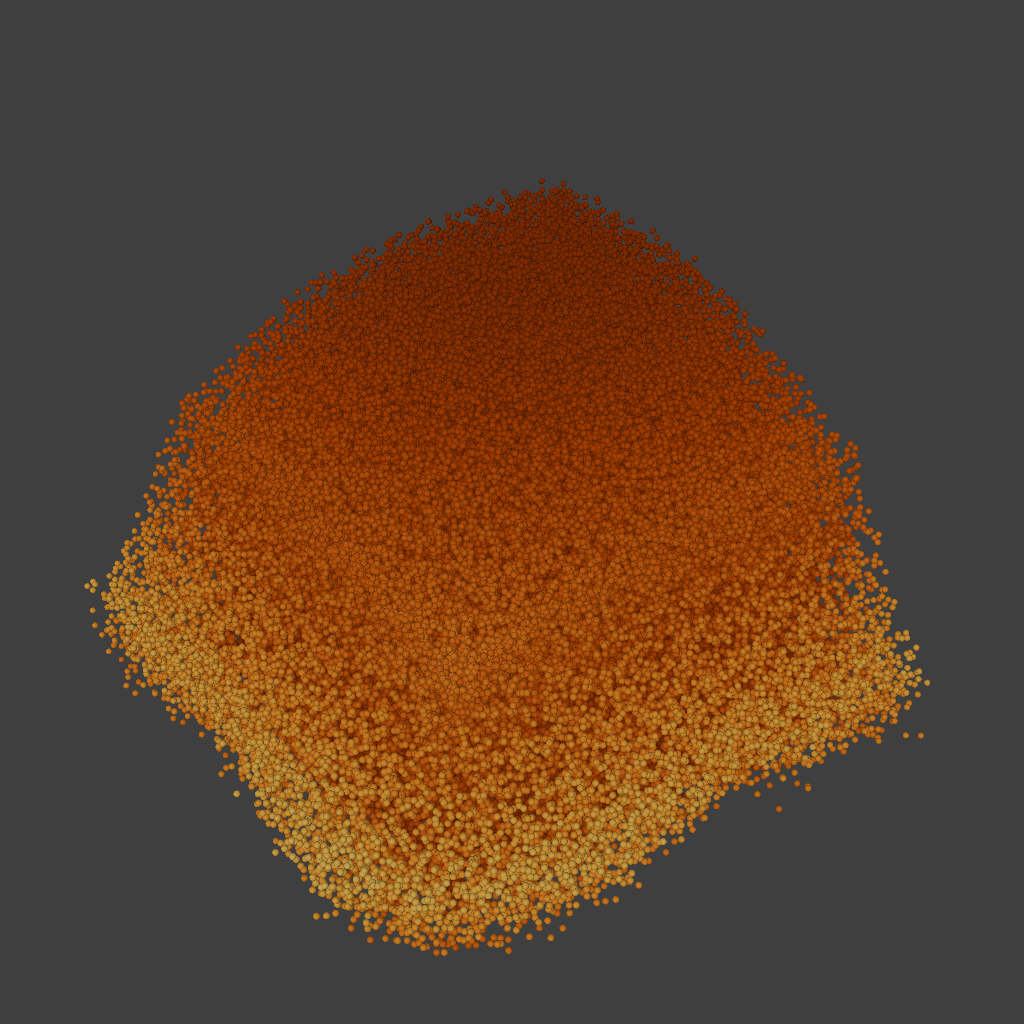}
    \caption{Rendered aerial view of in situ viz of 100k particles simulation with MPM and Galaxy at 5000 steps.}
    \label{fig:100k}
\end{figure}

\begin{figure}
    \centering
    \includegraphics[width=\linewidth]{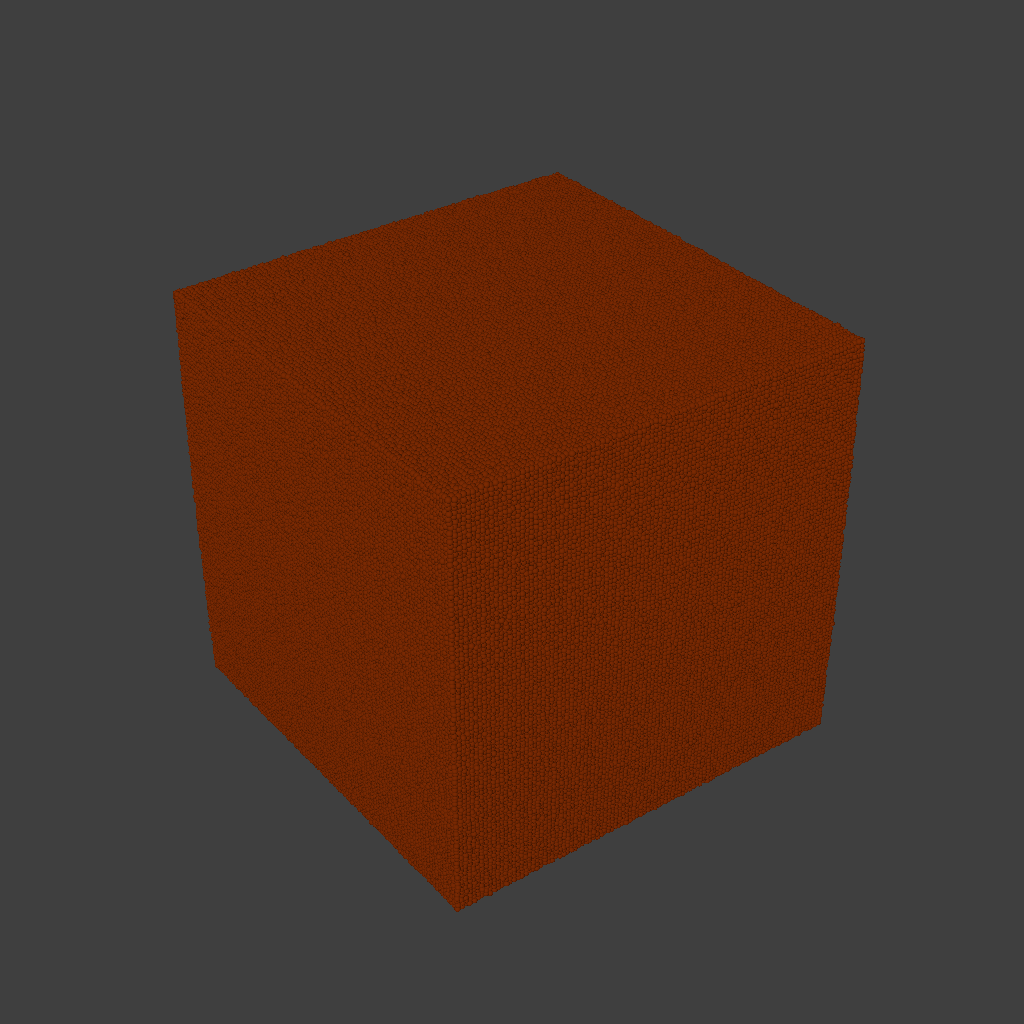}
    \caption{Rendered aerial view of in situ viz of 1M particles simulation with MPM and Galaxy at 550 steps.}
    \label{fig:1M}
\end{figure}

We simulate large-scale runs with 100,000 and 1 million material points. We use the same GNS predictions, as the problem domain remains the same, but we have increased the resolution of the MPM simulations by decreasing the particle size. For these large-scale runs, we exploit the MPI parallelization scheme with multiple nodes running the MPM simulations and the in situ render runs on a single process. Fig.~\ref{fig:100k} shows the aerial view of the 100k simulation with MPM and Galaxy. The initial stages of the collapse for the 1 million particles is shown in fig.~\ref{fig:1M}. 

Table~\ref{tab:totalruntimes} shows the average run time for 13k, 100k, and 1M particles in situ visualization with MPM and Galaxy. As the number of particles increases from 13k to 100k, the setup time remains unchanged (5.3\%); however, the amount of time required to transfer data increases by 24x, now consuming about 50\% of the total in situ viz time. The remaining 44\% is spent on rendering. We observe that the proportion of rendering time decreases as the problem size increase. For 1 million particles, the render time drops to only 18.7\%, while the data transfer consumes nearly 62\% of the total viz time. In these cases, we use a single in situ process to receive all the MPM particle data to render. Using multiple processes will optimize this data transfer time as it will be distributed across nodes. Galaxy supports distributed asynchronous rendering, thus reducing the transfer time and the total in situ viz time.

\begin{table}[h!]
    \centering
    \caption{In situ visualization time for each step of 13k, 100k and 1M particles with MPM and Galaxy.}
    \label{tab:totalruntimes}
    \begin{tabular}{l c c c} 
     \toprule
     Run & \multicolumn{3}{c}{Avg. run time / in situ viz step (s)} \\ 
     \cmidrule{2-4}
     & 13k & 100k & 1M \\
     \midrule
     Receive & 0.052 (23.2\%) & 1.281 (50.8\%) & 2.078 (61.6\%)\\
     Setup & 0.012  (5.3\%) & 0.133 (5.3\%)& 0.666 (19.7\%)\\
     Render & 0.16  (71.5\%)& 1.107 (43.9\%)& 0.632 (18.7\%)\\
     \midrule
     Total & 0.224  & 2.521 & 3.376 \\
     \bottomrule
     \end{tabular}
 \end{table}

\section{Future Work}
\label{sec:futurework}
\begin{figure}
    \centering
    \includegraphics[width=\linewidth]{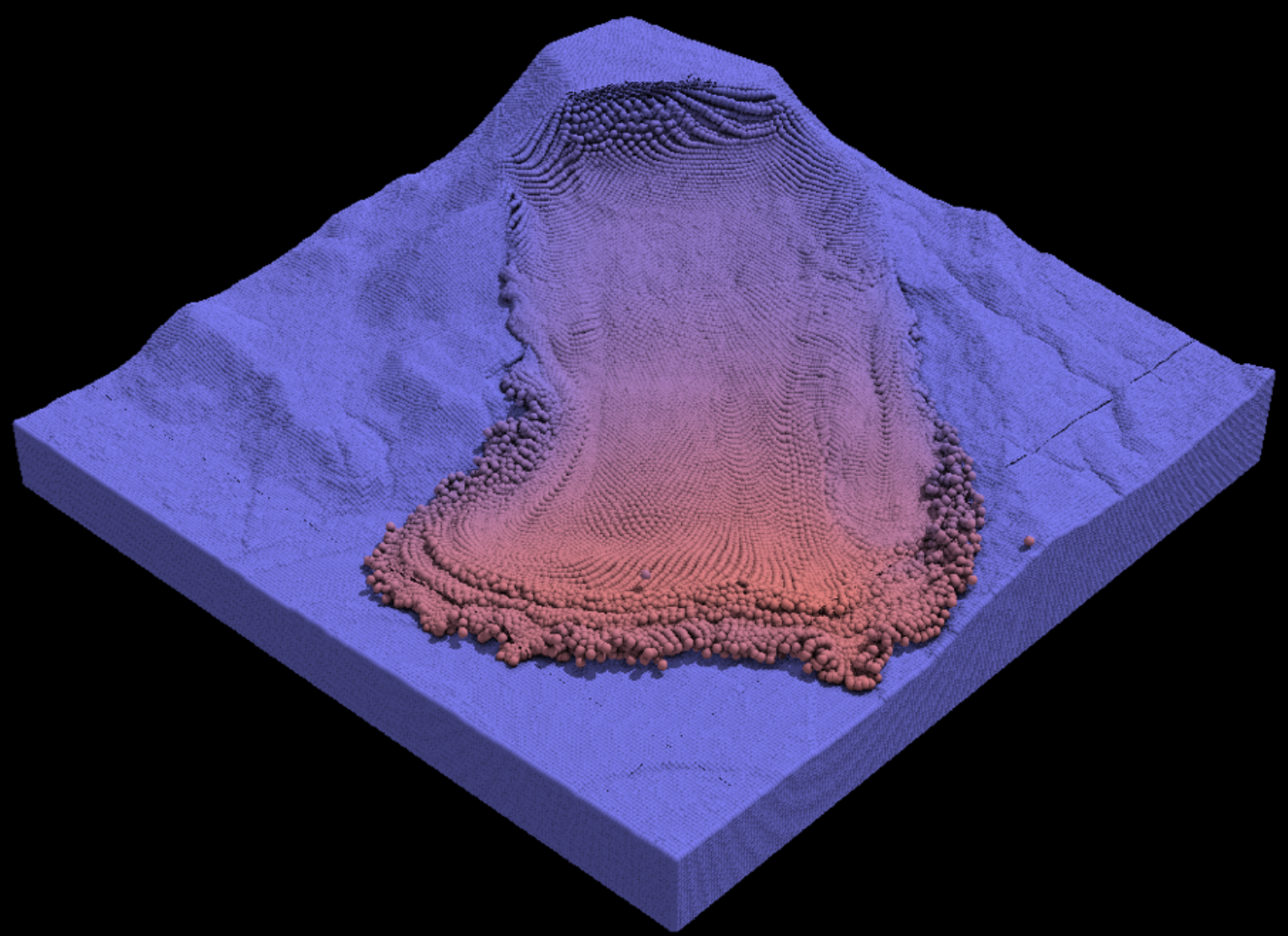}
    \caption{Distributed asynchronous rendering of the Oso landslide with TACC Galaxy and CB-Geo MPM.}
    \label{fig:oso}
\end{figure}

We provide an example simulation of granular column collapse, a classical granular physics problem that captures the dynamics of a landslide. We have scaled the GNS-informed in situ approach to run one million particles. A regional-scale landslide simulation would require 100s of millions to billions of particles. GNS is capable of predicting systems that are larger than the problem domains used in the training dataset. For GNS-informed visualization, we hope to run a coarse-scale GNS prediction of the landslide evolution and identify appropriate viewports, camera angles, and features. Abrams et~al.~\cite{abram22insitu} simulated a regional-scale runout of the Oso landslide with 5 million material points (see fig.~\ref{fig:oso}). However, this required several trials to obtain the correct view and parameters to render. Using the GNS-informed in situ visualization approach, we can pre-determine the correct viewports and rendering parameters. Using these preset configurations, we will run a distributed asynchronous ray tracing with Galaxy and multi-node parallel MPM simulation of the landslide. We will also utilize distributed data transfer running multiple rendering processes to minimize transfer time bottlenecks. The GNS-informed in situ approach offers the best of both worlds of in situ visualizations of peta- and exascale simulations while retaining the benefits of post hoc visualization through GNS predicted trajectories.

\section{Conclusion}
\label{sec:conclusion}
In situ visualization techniques are hampered by a lack of foresight: crucial simulation phenomena can be missed due to a poor sampling rate or insufficient detail at critical timesteps. We present GNS-informed in situ visualizations for large-deformation granular flow problems. We leverage the ML surrogate to pre-run the entire simulation of granular column collapse using GNS. We visualize the GNS results to identify critical regions, features, and camera angles for the in situ viz. We then use these metadata gathered from the GNS run to set up and simulate the real-scale in situ visualization of MPM with Galaxy. The GNS oracle successfully predicted the MPM dynamics observed in the full-scale simulations, including specific features tagged at different time steps and the displacement magnitudes. The GNS-informed approach reduces the number of trials and, in turn, the associated runtime costs required to capture the critical dynamics in an in situ visualization setting. By leveraging GNS surrogate, we offer a novel human-centered-approach to in situ viz. The GNS-informed approach offers the benefit of in situ visualizations, while maintaining the flexibility of a post hoc visualization.

\acknowledgments{
The authors would like to thank Dr. Greg Abram for fruitful discussions and help with the Galaxy set-up for in situ visualization. Krishna Kumar would like to thank the U.S. National Science Foundation grant CMMI-2022469 and OAC-2103937. This work was also funded in part by the Intel oneAPI Graphics and Visualization Institute of eXcellence at TACC.}

\bibliographystyle{abbrv-doi-hyperref}
\bibliography{gns4vis}

\end{document}